\documentclass[prl,amsmath,amssymb,twocolumn]{revtex4-2}
\usepackage{graphicx}
\usepackage{subfigure}
\usepackage{adjustbox}
\usepackage{bm}
\usepackage{bbm}
\usepackage{color}
\usepackage{braket}
\usepackage{standalone}
\usepackage{multirow}
\usepackage{tikz}
\usepackage{mathrsfs}
\usepackage{dsfont}
\usepackage[colorlinks,bookmarks=true,citecolor=blue,linkcolor=blue,urlcolor=blue]{hyperref}
\usepackage{cleveref}
\usepackage{comment}
\usepackage{mathtools}
\usepackage{soul}
\usepackage{orcidlink}

\begin{document}

\title{Filling-Sensitive Spectral Complexity from Hilbert-Space Holonomy in Fragmented Non-Hermitian Systems}

\author{Jiong-Hao Wang\orcidlink{0000-0002-4161-9614}}\thanks{jionghao.wang@fysik.su.se}
\author{Maria Zelenayova \orcidlink{0009-0004-4229-266X}}
\author{Christopher Ekman\orcidlink{0009-0003-6426-2376}}
\author{Emil J. Bergholtz \orcidlink{0000-0002-9739-2930}}
\thanks{emil.bergholtz@fysik.su.se}
\affiliation{Department of Physics, Stockholm University, AlbaNova University Center, 106 91 Stockholm, Sweden}

\date{\today}

\begin{abstract}

We show that Hilbert-space holonomy provides a geometric organizing principle for spectral reality in fragmented non-Hermitian many-body systems, complementary to conventional symmetry protection. In two minimal fragmented models, complex spectra can arise only within the most symmetric sectors: half filling in the fermion model and zero magnetization in the spin chain. Adding or removing a single particle, or flipping a single spin, renders the spectra entirely real despite unchanged periodic boundary conditions, reminiscent of boundary-condition sensitivity in systems with a non-Hermitian skin effect. We explain this by viewing nonreciprocal hopping amplitudes as a discrete gauge field on the Krylov graph: trivial holonomy permits a diagonal similarity transformation to the Hermitian limit, whereas nontrivial holonomy obstructs it and allows complex spectra. In certain regimes, trivial holonomy admits an emergent-boundary interpretation, and longer-range models exhibit finite real and complex regions governed by the same criterion.

\end{abstract}

\maketitle

\emph{Introduction.}--- 
Effective non-Hermitian Hamiltonians arise naturally in open, driven, monitored, and dissipative quantum systems~\cite{Ashida2020AP,Bergholtz2021RMP},
giving rise to intriguing phenomena absent in Hermitian systems~\cite{Ashida2020AP,Bergholtz2021RMP,Okuma2023ARCMP,Ding2022NRP,Xue2026PRL}.
The features intrinsic to non-Hermitian systems are deeply connected with each other. For example,
the non-Hermitian skin effect (NHSE)~\cite{Lee2016PRL,Xiong2018JPC,YaoWang2018PRL,Kunst2018PRL,Martine2018PRB,Lee2021PRB,Alsallom2022PRR,Gliozzi2024PRL,Shimomura2024PRL,Yoshida2024PRL,Hu2025PRL,Liu2024PRL} is usually accompanied by non-trivial topology of the complex energy spectrum~\cite{Okuma2020PRL,ZhangK2020PRL} and sensitivity of the spectrum to boundary conditions~\cite{Lee2016PRL,YaoWang2018PRL}.
Non-Hermitian physics has been extensively studied at the single-particle level~\cite{Lee2016PRL,Xiong2018JPC,YaoWang2018PRL,Kunst2018PRL,Martine2018PRB,Gong2018PRX,Kawabata2019PRX,Song2019PRL,Yokomizo2019PRL,wojcik2020prb,li2021prb,yang2024rpp,Li2022PRL,Cao2023PRL,Zhang2025arxivtop,Liang2022PRL,Zhao2025Nature},
and recently non-Hermitian many-body systems have attracted considerable attention~\cite{Lee2021PRB,Alsallom2022PRR,Gliozzi2024PRL,Shimomura2024PRL,Yoshida2024PRL,Hu2025PRL,Hamazaki2019PRL,Garcia2022PRX,Chen2023SP,Lee2020PRB,Yang2021PRL,Kawabata2022PRB,Faugno2022PRL,Meden2023RPP,Ekman2026arxiv,Shen2025NC,Zhang2025arxivmb}.
Another aspect that has drawn ample interest is the emergence of real spectra protected by discrete symmetries such as pseudo-Hermiticity and $\mathcal{PT}$ symmetry~\cite{Bender1998PRL,Mostafazadeh2002JMP,Bender2024RMP}. 
This raises the question of whether spectral reality can be organized by mechanisms complementary to conventional symmetries and motivates the search for many-body analogues of non-Hermitian spectral sensitivity, familiar from boundary-condition dependence and enhanced responses at exceptional points~\cite{Wiersig2014PRL,Hodaei2017Nature,Chen2017Nature,Budich2020PRL}.

In a seemingly orthogonal research direction, Hilbert space fragmentation~\cite{SalaPRX2020,Yang2020PRL,Pietracaprina2021,MoudgalyaPRX2022,Moudgalya2022,Adler2024Nature,Wang2025PRXQ}, where kinetic constraints cause the Hilbert space to be disconnected into isolated Krylov sectors, was discovered in the study of non-ergodicity and  non-thermalization~\cite{SalaPRX2020}.
Despite several recent works on Hilbert space fragmentation in the non-Hermitian context~\cite{Ghosh2024PRB,Shen2022CP,Wang2025CPL,Li2023PRR,Paszko2025},
the intersection of non-Hermitian physics and Hilbert space fragmentation remains largely unexplored. Notably, Ref.~\onlinecite{Ghosh2024PRB} established that real spectra can arise in strongly fragmented non-Hermitian systems in the limit of large interactions combined with global symmetry protection, thus suggesting that fragmentation and purely real spectra can share a common origin in dynamical constraints. However, the geometric mechanism underlying this connection, the conditions under which it holds across fillings, and the resulting phenomenology have remained open.

\begin{figure}[!t]
  \centering
  \includegraphics[width=0.48\textwidth]{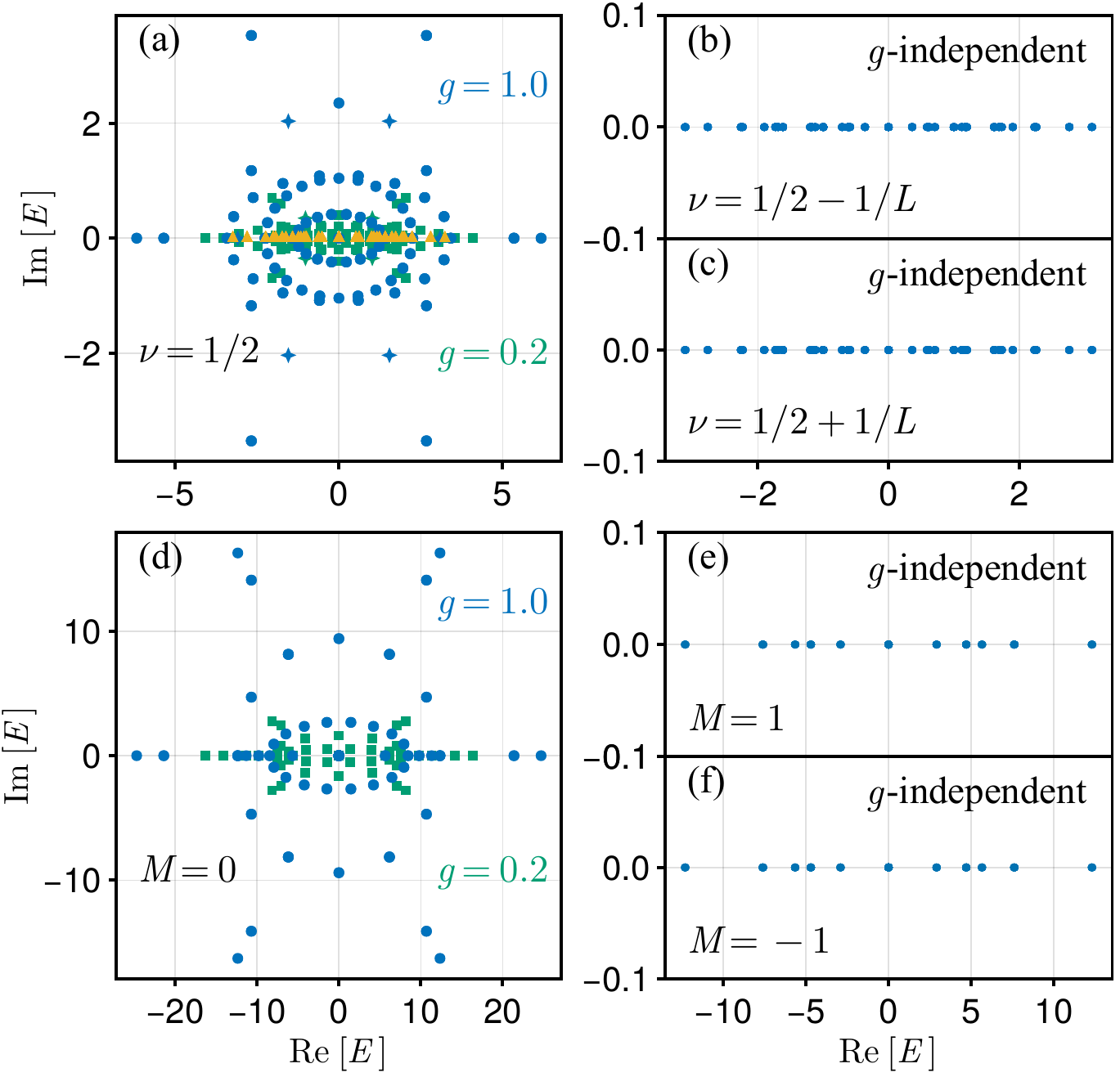}\hfill
  {\caption{\textbf{Filling-sensitive energy spectra}. (a)-(c) Energy spectra of the fermionic model with (\ref{Hop}) for system size $L=12$. $\nu=n_e/L$ is the filling factor with $n_e,L$ being the number of particles and sites, respectively. (a) Non-Hermitian case with $g=0.2$ (green squares) and $g=1.0$ (blue circles) at exact half-filling. The yellow triangles represent states with exactly the same energy for Hermitian and non-Hermitian cases and the stars indicate the states corresponding to the equivalent free fermion model with one fermion. (b) $g$-independent spectra with one particle removed from half-filling. (c) $g$-independent spectra with one particle added to half-filling.
  (d)-(f) Energy spectra for the spin model with (\ref{Spin}) for system size $L=6$. (d) Non-Hermitian case with $g=0.2$ (green squares) and $g=1.0$ (blue circles) at zero magnetization $M=0$. (e) $g$-independent spectra with $M=1$. (f) $g$-independent spectra with $M=-1$.
  }
  \label{fig_ComplexE}
  }
  
\end{figure}

In this Letter, we uncover a striking filling-sensitivity of the complex energy spectra in fragmented non-Hermitian many-body systems, and show that Hilbert-space holonomy provides an organizing principle for spectral complexity complementary to symmetry, with an intriguing connection to the (emergent) NHSE. Explicitly, complex energy spectra can only occur within the most symmetric symmetry sector, namely at half-filling for the minimal fermionic model [see Fig.~\ref{fig_ComplexE}(a)-(c)] and zero magnetization for the corresponding spin model [see Fig.~\ref{fig_ComplexE}(d)-(f)], and for all the other symmetry sectors, the spectra remain purely real.
In other words, a tiny doping away from the most symmetric sector---a single charge or hole to the fermionic system, or magnetized defect to the spin system---eliminates the spectral complexity completely. 
We show that the sharp filling-sensitivity of spectra originates from the rearrangement of the Hilbert-space structure upon changing the particle number or magnetization.
In the most symmetric sector, the non-trivial discrete holonomy of the Hilbert space obstructs this diagonal similarity transformation restoring the Hermitian limit and thus allows the complex spectra.
Adding or removing a single particle, or flipping a single spin, will lead to trivial holonomy enabling 
a diagonal similarity transformation, rendering Hermitian-like spectra even under periodic boundary conditions (PBCs).
Over broad filling ranges, trivial holonomy manifests as emergent open boundaries, 
with respect to which NHSE can occur without complex spectral winding.
More generally, fragmented non-Hermitian models with longer-range terms exhibit coexisting real and complex regimes, also governed by the holonomy criterion.

\emph{Minimal Models.}--- We consider two non-reciprocal models on a one-dimensional lattice with the unified form of Hamiltonian
\begin{equation}\label{Ham}
H(g)=\sum_{i} [e^g \hat{D}_i +e^{-g} \hat{D}_i^\dagger ].
\end{equation}
One is an interacting fermionic model with
\begin{equation}\label{Hop}
\hat{D}_i=c^\dagger_i c^\dagger_{i+3} c_{i+2}c_{i+1},\quad \hat{D}_i^\dagger=c^\dagger_{i+1} c^\dagger_{i+2} c_{i+3}c_{i},
\end{equation}
and another is a spin-1 chain with
\begin{equation}\label{Spin}
\hat{D}_i=S_i^+(S_{i+1}^-)^2S_{i+2}^+,\quad \hat{D}_i^\dagger=S_i^-(S_{i+1}^+)^2S_{i+2}^-.
\end{equation}
For the fermionic model with (\ref{Hop}), $\hat{D}_i$ and $\hat{D}_i^\dagger$ describe outward and inward hopping of two interacting particles, respectively.
In the Fock state representation, $\hat{D}_i$ renders the transition from (0110) to (1001), where 1 represents an occupied site and 0 an empty one, and $\hat{D}_i^\dagger$ vice versa.
For the spin model with (\ref{Spin}), $\hat{D}_i$ maps
$(-+-)\rightarrow(0-0)$, $(0+0)\rightarrow(+-+)$, $(-+0)\rightarrow(0-+)$ and $(0+-)\rightarrow(+-0)$, 
and $\hat{D}_i^\dagger$ backwards, where $-,0,+$ are eigenstates of the local spin operator $S^z$ with eigenvalues $-1,0,1$, respectively. 

The fermionic model conserves the particle number, and the spin model conserves the total magnetization $M=\sum_i S_i^z$.  In addition, both models conserve a dipole moment modulo $L$, so the Krylov decomposition occurs within symmetry sectors labeled by these global quantum numbers.
Both models are prototypical examples of Hilbert space fragmentation~\cite{SalaPRX2020,Moudgalya2020MV,Moudgalya2022,Bergholtz2005PRL}, where the dynamical constraints of the Hamiltonian disconnects the Hilbert space into an exponential number of isolated Krylov sectors.
Unlike the polynomially many symmetry sectors labeled by conserved quantum numbers, Krylov sectors cannot be distinguished by quantum numbers.

At half-filling $\nu=1/2$ of the fermionic model, within the specific Krylov sectors that every grouped two sites contain exactly one particle,
$H(g)$ on a lattice with length $L$ can be mapped to the non-interacting fermionic Hatano-Nelson (HN) model $H(g)=\sum_{\tilde{i}} [e^g d^\dagger_{\tilde{i}} d_{\tilde{i}+1} +e^{-g} d^\dagger_{\tilde{i}+1} d_{\tilde{i}} ]$ on an effective lattice with length $L/2$~\cite{Gliozzi2024PRL}.
Here, the operator $d^\dagger_{\tilde{i}}:=c_{2\tilde{i}-1}^\dagger c_{2\tilde{i}}$ creates a local dipole, introduced in the Hermitian limit in Ref.~\onlinecite{Bergholtz2005PRL}.
The mapping indicates complex energy spectra under PBCs, as shown in Fig.~\ref{fig_ComplexE}(a), and NHSE of dipoles~\cite{Gliozzi2024PRL}.
Here, in Fig.~\ref{fig_ComplexE}(a), the states represented by stars correspond to the equivalent free fermionic HN model with one fermion, exhibiting non-trivial spectral winding.
The model is also of interest at other filling factors in the Hermitian limit~\cite{Nakamura2012PRL},
and we might expect complex spectra for other fillings under PBCs as well.
All the discussions below are under PBCs unless stated otherwise.

\emph{Real spectra.}--- Remarkably, for the fermionic model, we find that at any filling factor away from exact half-filling the non-Hermitian energy spectrum is purely real and identical to the Hermitian case under PBCs, independent of $g$.
In contrast, at half-filling, complex spectra arise for nonzero $g$ and larger $g$ amplifies the maximal magnitude of eigenenergies [see Fig.~\ref{fig_ComplexE}(a)]. 
This suggests that the spectral complexity is sensitive to the total filling.
Specifically, in the thermodynamic limit, adding or removing a particle from half-filling, will render the spectra completely real, as shown in Fig.~\ref{fig_ComplexE}(a)-(c). 
Similarly, the spin model with (\ref{Spin}) only exhibits a complex spectrum at magnetization $M=0$. Once nonzero magnetization is turned on, no matter how small it is, Hermitian-like real spectra are restored, as shown in Fig.~\ref{fig_ComplexE}(d)-(f).

The spectral sensitivity to fillings and magnetization is rooted in the structure of the many-body Hilbert space, which we will show in the following.

\begin{figure}[!t]
  \centering
  \includegraphics[width=0.48\textwidth]{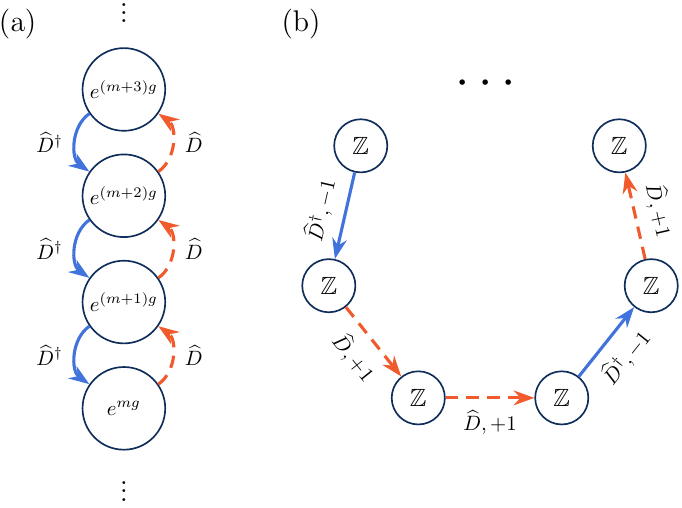}\hfill
  {\caption{\textbf{Schematic of the discrete holonomy}. The Hilbert space constitutes a topological graph, with basis states corresponding to vertices (indicated by circles) and $\hat{D}_i$ and $\hat{D}^\dagger_i$ corresponding to directed edges (indicated by arrows). (a) Schematic of the diagonal similarity transformation $\tilde{S}$. Each basis state is associated with a  matrix element of $\tilde{S}$ as labeled, with $\hat{D}_i$ and $\hat{D}^\dagger_i$ corresponding to adding and subtracting a $g$ in the exponent. 
  (b) Schematic of a closed loop in the space of basis states. Each state is associated with a $\mathbb{Z}$ bundle, with transports $\hat{D}_i$ and $\hat{D}^\dagger_i$ corresponding to +1 and -1, respectively, analogous to holonomy in differential geometry, as detailed in the End Matter.
  In the viewpoint of gauge potential, $\hat{D}_i$ and $\hat{D}^\dagger_i$ are interpreted as vector potential $A=+1$ and $-1$, and trivial and non-trivial holonomy are equivalent to $\phi=\sum_\mathcal{C}A=0$ or $\phi\neq0$.    
  }
  \label{fig_holo}
  }
  
\end{figure}

\emph{Discrete holonomy.}--- The identity of the energy spectra between Hermitian and non-Hermitian regime except at half-filling and zero magnetization implies a hidden similarity transformation under PBCs.
The intuitive conjecture is that Hamiltonian (\ref{Ham}) is related to its Hermitian counterpart by a diagonal similarity transformation $\tilde{S}$ under PBCs except at half-filling and zero magnetization, i.e., $\tilde{S}H(g)\tilde{S}^{-1}=H(0)$.
The matrix element of $\tilde{S}$ is $e^{m_jg}$ with integer $m_j$ for basis state $|f_j\rangle$, where
\begin{align}
m_j - m_k = 1
\quad \text{if} \quad
|f_k\rangle = \hat{D}_i |f_j\rangle,\\
m_j - m_k = -1
\quad \text{if} \quad
|f_k\rangle = \hat{D}_i^\dagger |f_j\rangle,
\end{align}
as illustrated in Fig.~\ref{fig_holo}(a).
We take direct product states as basis states $|f_j\rangle$, namely, Fock states for the fermionic system and sequences of $-,0,+$ for the spin system.
For a well-defined $\tilde{S}$, $|f_j\rangle$ and $m_j$ must have a one-to-one correspondence, which is dependent on the Hilbert space structure.

The Hilbert space structure is represented by a topological graph, where each Fock state corresponds to a vertex, and $\hat{D}_i$ and $\hat{D}^\dagger_i$ correspond to directed edges connecting the vertices, as shown in Fig.~\ref{fig_holo}.
The graphs corresponding to different Krylov sectors are isolated, therefore we call them Krylov graphs.
The graphs contain closed loops as shown in Fig.~\ref{fig_holo}(b), which represents a sequence of $\hat{D}_i$ and $\hat{D}^\dagger_i$ transporting a basis state $|f_j\rangle$ back to itself.
In loop $\mathcal{C}$, we denote the number of $\hat{D}_i$ and $\hat{D}^\dagger_i$ as $N_{D}$ and $N_{D^\dagger}$, respectively.
For $\tilde{S}$ to be well-defined we must require that 
every closed loop possesses an equal number of $\hat{D}_i$ and $\hat{D}^\dagger_i$, i.e.,
\begin{equation}\label{Identity}
N_{D} =N_{D^\dagger},\quad \forall\, \mathcal{C},
\end{equation}
which ensures that each $|f_j\rangle$ is uniquely associated with an $m_j$.
The condition Eq.~\ref{Identity} corresponds to the \textit{discrete holonomy} of the Hilbert space, analogous to the holonomy of the continuous manifold in differential geometry~\cite{Tu2017Book}, as detailed in the End Matter. 
If a sector of the Hilbert space satisfies Eq.~\ref{Identity}, we say that sector hosts trivial discrete holonomy.
Trivial holonomy enables the construction of the diagonal similarity transformation $\tilde{S}$, ensuring a real spectrum.
Otherwise, a loop with unequal numbers of $\hat{D}_i$ and $\hat{D}^\dagger_i$ defines non-trivial holonomy, which forbids the diagonal similarity transformation above and therefore allows a complex spectrum.
We can view trivial versus non-trivial holonomy as zero or nonzero flux through the loops.
In this perspective, directed edges representing $\hat{D}_i$ and $\hat{D}^\dagger_i$ are associated with a gauge potential $A=+1$ and $-1$, respectively, thus the flux through a loop $\mathcal{C}$ is $\phi=\sum_{\mathcal{C}}A$.
In a Hilbert space sector with trivial holonomy, the flux is zero for all the loops, while for non-trivial holonomy, there exists at least one loop with $\phi\neq 0$.

The minimal models can be shown to have trivial holonomy away from the most symmetric sectors
analytically by a height-function argument. We provide a sketch of the proof here with further details given in the Supplemental Material (SM)~\cite{SupMat}. For the fermionic chain, we define
\[
h_p=\sum_{i=1}^{p}(2n_i-1)-p\frac{2n_e-L}{L},
\qquad
V=\sum_{p=1}^{L}(h_p-\bar h)^2 ,
\]
where \(n_i=0,1\), \(n_e\) is the total particle number, and \(\bar h=L^{-1}\sum_p h_p\). The variance \(V\) is independent of the cyclic choice of origin. Under an inward move \(\hat D_i^\dagger:1001\to0110\), \(V\) changes by
\[
\Delta V_1=\frac{8(L-2n_e)}{L},
\]
whereas the reverse move \(\hat D_i\) changes \(V\) by \(-\Delta V_1\). Since \(V\) must return to its original value after any closed loop in a Krylov sector, a loop with \(N_D\) outward and \(N_{D^\dagger}\) inward moves obeys
\[
(N_{D^\dagger}-N_D)\Delta V_1=0 .
\]
Thus \(N_D=N_{D^\dagger}\) whenever \(n_e\neq L/2\), proving trivial holonomy in every Krylov component away from half filling.

The spin chain is analogous. Defining
\[
h_p=\sum_{i=1}^{p}S_i^z-p\frac{M}{L},
\qquad
V=\sum_{p=1}^{L}(h_p-\bar h)^2 ,
\]
one finds that a valid \(\hat D_i\) move changes \(V\) by \(2M/L\), while \(\hat D_i^\dagger\) changes it by \(-2M/L\). Hence every closed loop has equal numbers of \(\hat D_i\) and \(\hat D_i^\dagger\) moves for \(M\neq0\). All Krylov components away from zero magnetization therefore have trivial discrete holonomy and are diagonally similar to the Hermitian limit.

At half-filling for the fermionic model and zero magnetization for the spin model, the complex energy spectra in Fig.~\ref{fig_ComplexE}(a) and (d) suggest that the discrete holonomy of the Hilbert space is non-trivial. For example, for the fermionic model, in the Krylov sectors mappable to the non-interacting HN model~\cite{Gliozzi2024PRL},  we can form a loop with non-trivial holonomy by starting with, e.g., the state $|01100110\rangle$ and applying inward hopping $\hat{D}_i^\dagger$ only, in the following way:
\begin{equation}
\begin{aligned}
|01100110\rangle &\rightarrow |01011010\rangle
\rightarrow |10011001\rangle \\
&\rightarrow |01101001\rangle
\rightarrow |01100110\rangle.
\end{aligned}
\end{equation}
We see $N_{D^\dagger}=4$ and $N_{D} =0$, in violation of Eq.~\ref{Identity}, thus
demonstrating the non-trivial discrete holonomy. 
We show other examples of both models in the End Matter.
Note that there are also Krylov sectors with trivial discrete holonomy at half-filling, giving rise to real spectra identically to the Hermitian case [see the yellow triangles in Fig.~\ref{fig_ComplexE}(a)].

\begin{figure}[!t]
  \centering
  \includegraphics[width=0.48\textwidth]{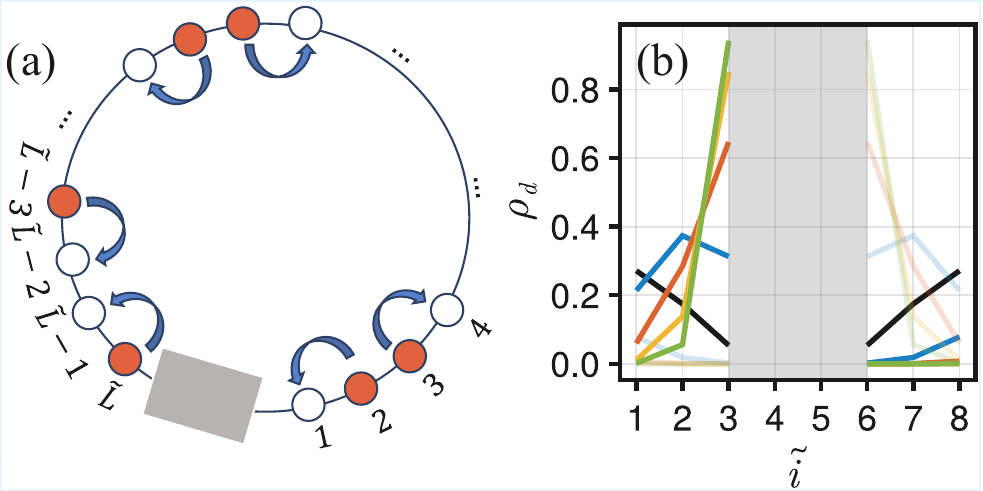}\hfill
  {\caption{\textbf{Emergent open boundaries and NHSE}. (a) Schematic of the EOBs. The gray box stands for the EOB, i.e., sites that no particles can hop across. $\tilde{L}$ is the number of sites except those in the EOBs. (b) Demonstration of the NHSE of dipoles with respect to the EOB by spatial distribution of the dipole density $\rho_d=\langle d^\dagger_{\tilde{i}} d_{\tilde{i}}\rangle$, with the dipole operator at composite site $\tilde{i}$ defined as $d^\dagger_{\tilde{i}}:=c_{2\tilde{i}-1}^\dagger c_{2\tilde{i}}$.
  Dipole density for right and left eigenvectors are shown by opaque and transparent lines, respectively, with the EOB indicated by gray.
  The results are obtained for the ground state of the Krylov sector containing $|\underline{10}\;\underline{01}\;\underline{01}\;\boxed{\underline{00}\;\underline{00}}\;\underline{01}\;\underline{01}\;\underline{01}\rangle$, with each composite site indicated by an underline and the EOB indicated by a box. The sequence apart from the EOB can be mapped to a free fermionic HN model with one fermion. 
  The results at $g=0.0,0.5,1.0,1.5$ and $2.0$ are indicated by black, blue, red, yellow and green colors, respectively, with $\rho_d$ for the right (left) eigenvector at site $\tilde{i}=3$ (6) increasing, demonstrating the NHSE.
  }
  \label{fig_OBC}
  }
  
\end{figure}

\emph{Emergent-boundary interpretation.}--- Over a range of filling factors for the fermionic model, the trivial discrete holonomy manifests itself in a simple physical picture of \textit{emergent open boundaries} (EOBs).  
For filling factors $\nu<1/3$ and $\nu>2/3$ (the two ranges are related by particle-hole symmetry transformation), we prove that every Krylov sector contains at least one invariant site~\cite{SupMat}, which plays the role of an EOB as no hopping can occur across this site.
Due to the EOBs, we can define a similarity transformation $SH(g)S^{-1}=H(0)$ with $S=\exp [-g \sum_{i=1}^L i^2 \hat{n}_{i}/4]$, similar to that under ordinary open boundary conditions (OBCs)~\cite{Gliozzi2024PRL}, by labeling the spatial position of the sites with respect to the EOBs, as schematically shown in Fig.~\ref{fig_OBC}(a), ensuring real spectra.
For a lattice under PBCs, if there are no EOBs, such a similarity transformation fails in general, as the spatial positions of the sites are ill-defined. 
For filling factors $1/3\leq\nu\leq 2/3$, although the EOB picture fails in general, there also exist certain Krylov sectors with EOBs, with an example shown in SM~\cite{SupMat}.

It is natural to ask whether NHSE can arise with respect to the EOBs, in analogy to ordinary OBCs.
We can take a Krylov sector in which each Fock state possesses patches mappable to the non-interacting HN model~\cite{Gliozzi2024PRL}, bounded by EOBs. 
For example, in the Fock state $|\underline{10}\;\underline{01}\;\underline{01}\;\underline{01}...\underline{01}\;\boxed{000}\rangle$,
the 0s in the box serve as an EOB, while the rest of the chain is split into pairs -- which we call composite sites -- as indicated by the underlines.
Each composite site indicated by a single underline contains one particle, therefore the patch excluding the EOB can be mapped to a free fermionic HN model,
with one $\underline{10}$ corresponding to one free fermion.  
In Fig.~\ref{fig_OBC}(b), we show the spatial distribution of the dipole density $\rho_{d}=\langle d^\dagger_{\tilde{i}} d_{\tilde{i}}\rangle$ under various non-Hermitian strength $g$.
We see that as $g$ increases, the right and left eigenstates, represented by opaque and transparent lines, respectively, accumulate at opposite sides of the EOB, indicated by gray.
Therefore, a NHSE of dipoles emerges with respect to the EOB,
along with purely real energy spectra. Thus skin localization appears within a Krylov sector whose spectrum remains real and $g$-independent. This differs from the usual HN setting, where skin localization is tied to the contrast between a winding PBC spectrum and an OBC spectrum.

\begin{figure}[!t]
  \centering
  \includegraphics[width=0.48\textwidth]{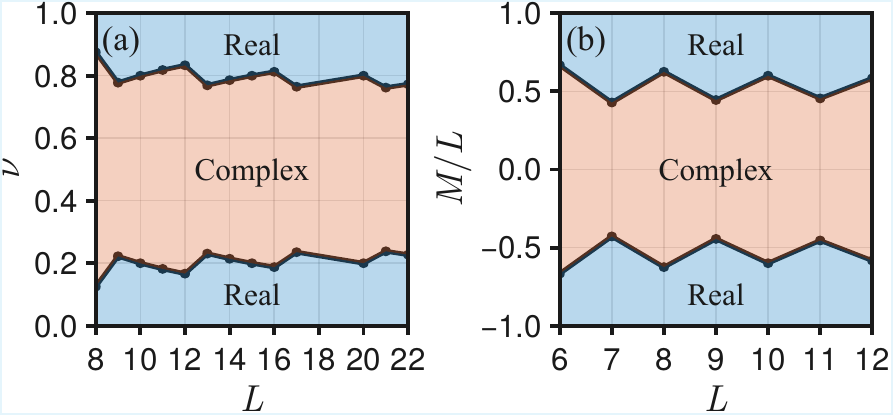}\hfill
  {\caption{\textbf{Phase diagram of longer-range models}. Phase diagram of complex and purely real spectra for (a) fermionic model with longer-range hopping (\ref{Long}) and (b) spin model with longer-range interaction (\ref{SpinLong}). We take $k_{\mathrm{max}}=3$ in (a). The phase diagrams are symmetric about $\nu=1/2$ in (a) and $M=0$ in (b) due to the symmetry. The filling factors and magnetization exactly on the phase boundaries (black lines) host real spectra. In (a), the largest (smallest) filling factor below (above) $\nu=1/2$ with real spectra is exactly $\nu=1/5$ when $L$ is an integer multiple of $5$. In (b), The minimal magnitude of $M$ with real spectra is $|M|=(L-1)/2$ for odd $L$ and $|M|=L/2+1$ for even $L$.   
  }
  \label{fig_longrange}
  }
  \end{figure}

\emph{Longer-range models.}--- For generality, we also investigate the fermionic model with longer-range hoppings
\begin{align}\label{Long}
H_\mathrm{long}^\mathrm{fermion}(g)=\sum_{i}&\sum_{0<l<k}^{k\leq k_\mathrm{max}}  \big[ 
 e^{klg/2} \, c^\dagger_{i} c^\dagger_{i+k+l} c_{i+k} c_{i+l} \nonumber\\
+\, & e^{-klg/2} \, c^\dagger_{i+l} c^\dagger_{i+k} c_{i+k+l} c_{i} 
\big],
\end{align} 
and the spin-1 model with a longer-range interaction
\begin{equation}\label{SpinLong}
\begin{aligned}
H_\mathrm{long}^\mathrm{spin}(g)=\sum_{i}\big[&e^{g}S_i^+S_{i+1}^-S_{i+2}^-S_{i+3}^+ \\+& e^{-g}S_i^-S_{i+1}^+S_{i+2}^+S_{i+3}^-
\big],
\end{aligned}
\end{equation}
which are also fragmented~\cite{SalaPRX2020,Moudgalya2022}.
In the phase diagram Fig.~\ref{fig_longrange}, we see finite regions of both real and complex spectra.
The Hilbert spaces host trivial discrete holonomy in the real regimes and non-trivial holonomy in the complex regimes, as detailed in the End Matter, which we have confirmed by computer programs up to system sizes shown in Fig.~\ref{fig_longrange}.

Hilbert space fragmentation is classified into strong and weak versions~\cite{SalaPRX2020}, depending on the absence or presence of a dominant Krylov sector. In the End Matter, we show trivial holonomy and real spectra can occur across strong and weak fragmentation, underscoring holonomy as a general principle for real spectra in non-Hermitian systems with non-reciprocity.

\emph{Conclusion.}--- In summary, we find sharp filling- and magnetization-sensitivity of spectral complexity in minimal fragmented non-Hermitian many-body systems, that is, complex energy spectra under PBCs only arise at the most symmetric symmetry sector: exactly half-filling for the fermionic model and zero magnetization for the spin model. 
We show that the sensitivity is rooted in the rearrangement of the Hilbert space structure upon adding or removing a particle, or flipping a spin.
Away from half-filling and zero magnetization, the Hilbert space has trivial discrete holonomy, enabling a diagonal similarity transformation to the Hermitian limit and restoring real spectra.
Otherwise, in the most symmetric symmetry sector, non-trivial holonomy forbids diagonal similarity transformations and therefore allows complex spectra.
Over a range of fillings, the trivial holonomy manifests itself in emergent open boundaries, with respect to which NHSE can arise without complex spectral winding. 
Fragmented non-Hermitian models with longer-range terms exhibit finite ranges of fillings and magnetization with real and complex spectra, governed by the same holonomy criterion.
Our work shows that holonomy provides a general organizing principle for spectral reality in non-reciprocal systems.

The filling sensitivity found here provides a many-body analogue of the boundary-condition sensitivity in non-Hermitian systems. In the Hatano–Nelson chain, changing from periodic to open boundaries removes spectral winding. Here the physical boundary condition is unchanged, but changing particle number or magnetization rearranges the Krylov graph and removes the holonomy responsible for complex eigenvalues. A microscopic change in a global quantum number can therefore have a genuinely non-Hermitian macroscopic spectral consequence.

Our work opens a new avenue to the study of non-Hermitian many-body systems,
motivating intriguing directions for future research.
First, for Hilbert spaces with non-trivial holonomy, it is unclear whether complex energy spectra always arise or in some cases other mechanisms -- such as a  a similarity transformation with off-diagonal elements -- can also protect the real spectra.
Second, it is possible that our results are generalizable to non-fragmented systems, in other words, 
whether we can construct a physically relevant quantum many-body model where the Hilbert space is not fragmented but still has trivial holonomy. 
Third, the holonomy structure may provide a new perspective for the study of ergodicity and thermalization in fragmented many-body systems.
Finally, the holonomy of the Hilbert space is mapped to the structure of a topological graph, 
which might inspire further applications of graph-theoretic methods to constrained non-Hermitian dynamics.

\emph{Acknowledgments}---%
We thank Zhenghui Li for helpful discussions.
This work was supported by the Knut and Alice Wallenberg Foundation (2023.0256) and the Göran Gustafsson Foundation for Research in Natural Sciences and Medicine.

\bibliography{EOBC_ref.bib}

@article{Ashida2020AP,
  title = {Non-{H}ermitian Physics},
  author = {Ashida, Yuto and Gong, Zongping and Ueda, Masahito},
  year = 2020,
  month = jul,
  journal = {Adv. Phys.},
  volume = {69},
  number = {3},
  pages = {249--435},
  publisher = {Taylor \& Francis},
  issn = {0001-8732},
  doi = {10.1080/00018732.2021.1876991}
}

@article{Bergholtz2021RMP,
  title = {Exceptional Topology of Non-{H}ermitian Systems},
  author = {Bergholtz, Emil J. and Budich, Jan Carl and Kunst, Flore K.},
  year = 2021,
  month = feb,
  journal = {Rev. Mod. Phys.},
  volume = {93},
  number = {1},
  pages = {015005},
  publisher = {American Physical Society},
  doi = {10.1103/RevModPhys.93.015005}
}

@article{Okuma2023ARCMP,
  title = {Non-{H}ermitian Topological Phenomena: A Review},
  shorttitle = {Non-Hermitian Topological Phenomena},
  author = {Okuma, Nobuyuki and Sato, Masatoshi},
  year = 2023,
  month = mar,
  journal = {Annu. Rev. Condens. Matter Phys},
  volume = {14},
  number = {1},
  pages = {83--107},
  issn = {1947-5454, 1947-5462},
  doi = {10.1146/annurev-conmatphys-040521-033133},
  copyright = {http://creativecommons.org/licenses/by/4.0/}
}

@article{Ding2022NRP,
  title = {Non-{H}ermitian Topology and Exceptional-Point Geometries},
  author = {Ding, Kun and Fang, Chen and Ma, Guancong},
  year = 2022,
  journal = {Nat. Rev. Phys.},
  volume = {4},
  number = {12},
  pages = {745--760},
  doi={10.1038/s42254-022-00516-5},
  publisher = {Nature Publishing Group UK London}
}

@article{Xue2026PRL,
  title = {Essay: Topological Phases and Exceptional Points in Non-{H}ermitian Systems},
  shorttitle = {Essay},
  author = {Xue, Peng},
  year = 2026,
  month = apr,
  journal = {Phys. Rev. Lett.},
  volume = {136},
  number = {17},
  pages = {170001},
  publisher = {American Physical Society},
  doi = {10.1103/ll76-j2l5}
}

@article{Lee2016PRL,
  title={Anomalous Edge State in a Non-{H}ermitian Lattice},
  author={Lee, Tony E.},
  journal={Phys. Rev. Lett.},
  volume={116},
  pages={133903},
  year={2016},
  doi={10.1103/PhysRevLett.116.133903}
}

@article{Xiong2018JPC,
  title = {Why Does Bulk Boundary Correspondence Fail in Some Non-{H}ermitian Topological Models},
  author = {Xiong, Ye},
  year = 2018,
  month = mar,
  journal = {J. Phys. Comm.},
  volume = {2},
  number = {3},
  pages = {035043},
  publisher = {IOP Publishing},
  issn = {2399-6528},
  doi = {10.1088/2399-6528/aab64a}
}

@article{YaoWang2018PRL,
  title={Edge States and Topological Invariants of Non-{H}ermitian Systems},
  author={Yao, Shunyu and Wang, Zhong},
  journal={Phys. Rev. Lett.},
  volume={121},
  pages={086803},
  year={2018},
  doi={10.1103/PhysRevLett.121.086803}
}

@article{Kunst2018PRL,
  title={Biorthogonal Bulk–Boundary Correspondence in Non-{H}ermitian Systems},
  author={Kunst, Flore K. and Edvardsson, Elias and Budich, Jan C. and Bergholtz, Emil J.},
  journal={Phys. Rev. Lett.},
  volume={121},
  pages={026808},
  year={2018},
  doi={10.1103/PhysRevLett.121.026808}
}

@article{Martine2018PRB,
  title = {Non-{H}ermitian Robust Edge States in One Dimension: Anomalous Localization and Eigenspace Condensation at Exceptional Points},
  shorttitle = {Non-{H}ermitian Robust Edge States in One Dimension},
  author = {Martinez Alvarez, V. M. and Barrios Vargas, J. E. and Foa Torres, L. E. F.},
  year = 2018,
  month = mar,
  journal = {Phys. Rev. B},
  volume = {97},
  number = {12},
  pages = {121401},
  publisher = {American Physical Society},
  doi = {10.1103/PhysRevB.97.121401}
}

@article{Okuma2020PRL,
  title={Topological Origin of Non-{H}ermitian Skin Effect},
  author={Okuma, Nobuyuki and Kawabata, Kohei and Shiozaki, Ken and Sato, Masatoshi},
  journal={Phys. Rev. Lett.},
  volume={124},
  pages={086801},
  year={2020},
  doi={10.1103/PhysRevLett.124.086801}
}

@article{ZhangK2020PRL,
  title = {Correspondence between Winding Numbers and Skin Modes in Non-{H}ermitian Systems},
  author = {Zhang, Kai and Yang, Zhesen and Fang, Chen},
  year = 2020,
  month = sep,
  journal = {Phys. Rev. Lett.},
  volume = {125},
  number = {12},
  pages = {126402},
  publisher = {American Physical Society},
  doi = {10.1103/PhysRevLett.125.126402}
}

@article{Lee2021PRB,
  title = {Many-Body Topological and Skin States without Open Boundaries},
  author = {Lee, Ching Hua},
  year = 2021,
  month = nov,
  journal = {Phys. Rev. B},
  volume = {104},
  number = {19},
  pages = {195102},
  issn = {2469-9950, 2469-9969},
  doi = {10.1103/PhysRevB.104.195102}
}

@article{Alsallom2022PRR,
  title = {Fate of the Non-{H}ermitian Skin Effect in Many-Body Fermionic Systems},
  author = {Alsallom, Faisal and Herviou, Lo{\"i}c and Yazyev, Oleg V. and Brzezi{\'n}ska, Marta},
  year = 2022,
  month = aug,
  journal = {Phys. Rev. Research},
  volume = {4},
  number = {3},
  pages = {033122},
  issn = {2643-1564},
  doi = {10.1103/PhysRevResearch.4.033122}
}

@article{Gliozzi2024PRL,
  title = {Many-Body Non-{H}ermitian Skin Effect for Multipoles},
  author = {Gliozzi, Jacopo and De Tomasi, Giuseppe and Hughes, Taylor L.},
  year = 2024,
  month = sep,
  journal = {Phys. Rev. Lett.},
  volume = {133},
  number = {13},
  pages = {136503},
  issn = {0031-9007, 1079-7114},
  doi = {10.1103/PhysRevLett.133.136503}
}

@article{Shimomura2024PRL,
  title = {General Criterion for Non-{H}ermitian Skin Effects and Application: {F}ock Space Skin Effects in Many-Body Systems},
  shorttitle = {General Criterion for Non-Hermitian Skin Effects and Application},
  author = {Shimomura, Kenji and Sato, Masatoshi},
  year = 2024,
  month = sep,
  journal = {Phys. Rev. Lett.},
  volume = {133},
  number = {13},
  pages = {136502},
  issn = {0031-9007, 1079-7114},
  doi = {10.1103/PhysRevLett.133.136502}
}

@article{Yoshida2024PRL,
  title = {Non-{H}ermitian {M}ott Skin Effect},
  author = {Yoshida, Tsuneya and Zhang, Song-Bo and Neupert, Titus and Kawakami, Norio},
  year = 2024,
  month = aug,
  journal = {Phys. Rev. Lett.},
  volume = {133},
  number = {7},
  pages = {076502},
  publisher = {American Physical Society},
  doi = {10.1103/PhysRevLett.133.076502}
}

@article{Liu2024PRL,
  title = {Dynamical Transition Due to Feedback-Induced Skin Effect},
  author = {Liu, Ze-Chuan and Li, Kai and Xu, Yong},
  year = 2024,
  month = aug,
  journal = {Phys. Rev. Lett.},
  volume = {133},
  number = {9},
  pages = {090401},
  publisher = {American Physical Society},
  doi = {10.1103/PhysRevLett.133.090401}
}

@article{Hu2025PRL,
  title = {Many-Body Non-{H}ermitian Skin Effect with Exact Steady States in the Dissipative Quantum Link Model},
  author = {Hu, Yu-Min and Wang, Zijian and Lian, Biao and Wang, Zhong},
  year = 2025,
  month = dec,
  journal = {Phys. Rev. Lett.},
  volume = {135},
  number = {26},
  pages = {260401},
  publisher = {American Physical Society},
  doi = {10.1103/wztw-l8wg}
}

@misc{Ekman2026arxiv,
  title = {Symmetry-Fractionalized Skin Effects in Non-{H}ermitian {L}uttinger Liquids},
  author = {Ekman, Christopher and Bergholtz, Emil J. and Molignini, Paolo},
  year = 2026,
  month = apr,
  number = {arXiv:2603.28849},
  eprint = {2603.28849},
  primaryclass = {cond-mat},
  publisher = {arXiv},
  doi = {10.48550/arXiv.2603.28849},
  archiveprefix = {arXiv}
}

@article{Gong2018PRX,
  title = {Topological Phases of Non-{H}ermitian Systems},
  author = {Gong, Zongping and Ashida, Yuto and Kawabata, Kohei and Takasan, Kazuaki and Higashikawa, Sho and Ueda, Masahito},
  year = 2018,
  month = sep,
  journal = {Phys. Rev. X},
  volume = {8},
  number = {3},
  pages = {031079},
  publisher = {American Physical Society},
  doi = {10.1103/PhysRevX.8.031079}
}

@article{Kawabata2019PRX,
  title={Symmetry and Topology in Non-{H}ermitian Physics},
  author={Kawabata, Kohei and Shiozaki, Ken and Ueda, Masahito and Sato, Masatoshi},
  journal={Phys. Rev. X},
  volume={9},
  pages={041015},
  year={2019},
  doi={10.1103/PhysRevX.9.041015}
}

@article{Song2019PRL,
  title = {Non-{H}ermitian Topological Invariants in Real Space},
  author = {Song, Fei and Yao, Shunyu and Wang, Zhong},
  year = 2019,
  month = dec,
  journal = {Phys. Rev. Lett.},
  volume = {123},
  number = {24},
  pages = {246801},
  publisher = {American Physical Society},
  doi = {10.1103/PhysRevLett.123.246801}
}

@article{Yokomizo2019PRL,
  title={Non-{B}loch Band Theory of Non-{H}ermitian Systems},
  author={Yokomizo, Katsunori and Murakami, Shuichi},
  journal={Phys. Rev. Lett.},
  volume={123},
  pages={066404},
  year={2019},
  doi={10.1103/PhysRevLett.123.066404}
}

@article{wojcik2020prb,
  ids = {wojcikHomotopyCharacterizationNonHermitian2020a},
  title = {Homotopy characterization of non-{H}ermitian {H}amiltonians},
  author = {Wojcik, Charles C. and Sun, Xiao-Qi and Bzdušek, Tomáš and Fan, Shanhui},
  date = {2020-05-15},
  journal = {Phys. Rev. B},
  year = {2020},
  volume = {101},
  number = {20},
  pages = {205417},
  publisher = {American Physical Society},
  doi = {10.1103/PhysRevB.101.205417}
}

@article{li2021prb,
  ids = {liHomotopicalCharacterizationNonHermitian2021a},
  title = {Homotopical Characterization of Non-{{Hermitian}} Band Structures},
  author = {Li, Zhi and Mong, Roger S. K.},
  date = {2021-04-16},
  journal = {Phys. Rev. B},
  year = {2021},
  volume = {103},
  number = {15},
  pages = {155129},
  publisher = {American Physical Society},
  doi = {10.1103/PhysRevB.103.155129}
}

@article{yang2024rpp,
  ids = {yangHomotopySymmetryNonHermitian2023,yangHomotopySymmetryNonHermitian2024a},
  title = {Homotopy, {{Symmetry}}, and Non-{{Hermitian Band Topology}}},
  author = {Yang, Kang and Li, Zhi and König, J. Lukas K. and Rødland, Lukas and Stålhammar, Marcus and Bergholtz, Emil J.},
  date = {2024-07},
  journaltitle = {Reports on Progress in Physics},
  journal = {Rep. Prog. Phys.},
  year = {2024},
  volume = {87},
  number = {7},
  pages = {078002},
  publisher = {IOP Publishing},
  issn = {0034-4885},
  doi = {10.1088/1361-6633/ad4e64},
  langid = {english}
}

@article{Li2022PRL,
  title = {Non-{H}ermitian Absorption Spectroscopy},
  author = {Li, Kai and Xu, Yong},
  year = 2022,
  month = aug,
  journal = {Phys. Rev. Lett.},
  volume = {129},
  number = {9},
  pages = {093001},
  publisher = {American Physical Society},
  doi = {10.1103/PhysRevLett.129.093001}
}

@article{Cao2023PRL,
  title = {Probing Complex-Energy Topology via Non-{H}ermitian Absorption Spectroscopy in a Trapped Ion Simulator},
  author = {Cao, M.-M. and Li, K. and Zhao, W.-D. and Guo, W.-X. and Qi, B.-X. and Chang, X.-Y. and Zhou, Z.-C. and Xu, Y. and Duan, L.-M.},
  year = 2023,
  month = apr,
  journal = {Phys. Rev. Lett.},
  volume = {130},
  number = {16},
  pages = {163001},
  publisher = {American Physical Society},
  doi = {10.1103/PhysRevLett.130.163001}
}

@misc{Zhang2025arxivtop,
  title = {Observation of Non-{H}ermitian Topology in Cold {R}ydberg Quantum Gases},
  author = {Zhang, Jun and Wang, Ya-Jun and Shao, Shi-Yao and Liu, Bang and Zhang, Li-Hua and Zhang, Zheng-Yuan and Liu, Xin and Yu, Chao and Li, Qing and Chen, Han-Chao and Ma, Yu and Han, Tian-Yu and Wang, Qi-Feng and Nan, Jia-Dou and Yin, Yi-Ming and Zhu, Dong-Yang and Fang, Qiao-Qiao and Ding, Dong-Sheng and Shi, Bao-Sen},
  year = 2025,
  month = sep,
  number = {arXiv:2509.26256},
  eprint = {2509.26256},
  primaryclass = {cond-mat},
  publisher = {arXiv},
  doi = {10.48550/arXiv.2509.26256},
  archiveprefix = {arXiv}
}

@article{Liang2022PRL,
  title = {Dynamic Signatures of Non-{H}ermitian Skin Effect and Topology in Ultracold Atoms},
  author = {Liang, Qian and Xie, Dizhou and Dong, Zhaoli and Li, Haowei and Li, Hang and Gadway, Bryce and Yi, Wei and Yan, Bo},
  year = 2022,
  month = aug,
  journal = {Phys. Rev. Lett.},
  volume = {129},
  number = {7},
  pages = {070401},
  publisher = {American Physical Society},
  doi = {10.1103/PhysRevLett.129.070401}
}

@article{Zhao2025Nature,
  title = {Two-Dimensional Non-{H}ermitian Skin Effect in an Ultracold {F}ermi Gas},
  author = {Zhao, Entong and Wang, Zhiyuan and He, Chengdong and Poon, Ting Fung Jeffrey and Pak, Ka Kwan and Liu, Yu-Jun and Ren, Peng and Liu, Xiong-Jun and Jo, Gyu-Boong},
  year = 2025,
  month = jan,
  journal = {Nature},
  volume = {637},
  number = {8046},
  pages = {565--573},
  publisher = {Nature Publishing Group},
  issn = {1476-4687},
  doi = {10.1038/s41586-024-08347-3},
  copyright = {2025 The Author(s), under exclusive licence to Springer Nature Limited}
}

@article{Hamazaki2019PRL,
  title = {Non-{H}ermitian Many-Body Localization},
  author = {Hamazaki, Ryusuke and Kawabata, Kohei and Ueda, Masahito},
  year = 2019,
  month = aug,
  journal = {Phys. Rev. Lett.},
  volume = {123},
  number = {9},
  pages = {090603},
  issn = {0031-9007, 1079-7114},
  doi = {10.1103/PhysRevLett.123.090603}
}

@article{Garcia2022PRX,
  title = {Symmetry Classification and Universality in Non-{H}ermitian Many-Body Quantum Chaos by the {Sachdev-Ye-Kitaev} Model},
  author = {{Garc{\'i}a-Garc{\'i}a}, Antonio M. and S{\'a}, Lucas and Verbaarschot, Jacobus J. M.},
  year = 2022,
  month = may,
  journal = {Phys. Rev. X},
  volume = {12},
  number = {2},
  pages = {021040},
  publisher = {American Physical Society},
  doi = {10.1103/PhysRevX.12.021040}
}

@article{Chen2023SP,
  title = {Weak Ergodicity Breaking in Non-{H}ermitian Many-Body Systems},
  author = {Chen, Qianqian and Chen, Shuai A. and Zhu, Zheng},
  year = 2023,
  month = aug,
  journal = {SciPost Phys.},
  volume = {15},
  number = {2},
  pages = {052},
  issn = {2542-4653},
  doi = {10.21468/SciPostPhys.15.2.052}
}

@article{Lee2020PRB,
  title = {Many-Body Approach to Non-{H}ermitian Physics in Fermionic Systems},
  author = {Lee, Eunwoo and Lee, Hyunjik and Yang, Bohm-Jung},
  year = 2020,
  month = mar,
  journal = {Phys. Rev. B},
  volume = {101},
  number = {12},
  pages = {121109},
  issn = {2469-9950, 2469-9969},
  doi = {10.1103/PhysRevB.101.121109}
}

@article{Yang2021PRL,
  title = {Exceptional Spin Liquids from Couplings to the Environment},
  author = {Yang, Kang and Morampudi, Siddhardh C. and Bergholtz, Emil J.},
  journal = {Phys. Rev. Lett.},
  volume = {126},
  issue = {7},
  pages = {077201},
  numpages = {7},
  year = {2021},
  month = {Feb},
  publisher = {American Physical Society},
  doi = {10.1103/PhysRevLett.126.077201},
  url = {https://link.aps.org/doi/10.1103/PhysRevLett.126.077201}
}

@article{Kawabata2022PRB,
  title = {Many-Body Topology of Non-{H}ermitian Systems},
  author = {Kawabata, Kohei and Shiozaki, Ken and Ryu, Shinsei},
  year = 2022,
  month = apr,
  journal = {Phys. Rev. B},
  volume = {105},
  number = {16},
  pages = {165137},
  issn = {2469-9950, 2469-9969},
  doi = {10.1103/PhysRevB.105.165137}
}

@article{Faugno2022PRL,
  title = {Interaction-Induced Non-{H}ermitian Topological Phases from a Dynamical Gauge Field},
  author = {Faugno, W. N. and Ozawa, Tomoki},
  year = 2022,
  month = oct,
  journal = {Phys. Rev. Lett.},
  volume = {129},
  number = {18},
  pages = {180401},
  publisher = {American Physical Society},
  doi = {10.1103/PhysRevLett.129.180401}
}

@article{Meden2023RPP,
  title = {$\mathcal{PT}$-Symmetric, Non-{H}ermitian Quantum Many-Body Physics---a Methodological Perspective},
  author = {Meden, Volker and Grunwald, Lukas and Kennes, Dante M.},
  year = 2023,
  journal = {Rep. Prog. Phys.},
  volume = {86},
  number = {12},
  pages = {124501},
  doi={10.1088/1361-6633/ad05f3},
  publisher = {IOP Publishing}
}

@article{Shen2025NC,
  title = {Observation of the Non-{H}ermitian Skin Effect and {F}ermi Skin on a Digital Quantum Computer},
  author = {Shen, Ruizhe and Chen, Tianqi and Yang, Bo and Lee, Ching Hua},
  year = 2025,
  month = feb,
  journal = {Nat. Comm.},
  volume = {16},
  number = {1},
  pages = {1340},
  publisher = {Nature Publishing Group},
  issn = {2041-1723},
  doi = {10.1038/s41467-025-55953-4},
  copyright = {2025 The Author(s)}
}

@misc{Zhang2025arxivmb,
  title = {Observation of Non-{H}ermitian Many-Body Phase Transition in a {R}ydberg-Atom Array},
  author = {Zhang, Yao-Wen and Xu, Biao and Zhou, Yijia and Xiang, De-Sheng and Liu, Hao-Xiang and Zhou, Peng and Zhang, Kuan and Liao, Ren and Pohl, Thomas and Li, Weibin and Li, Lin},
  year = 2025,
  month = dec,
  number = {arXiv:2512.02753},
  eprint = {2512.02753},
  primaryclass = {quant-ph},
  publisher = {arXiv},
  doi = {10.48550/arXiv.2512.02753},
  archiveprefix = {arXiv}
}

@article{Ghosh2024PRB,
  title = {Hilbert Space Fragmentation Imposed Real Spectrum of Non-{H}ermitian Systems},
  author = {Ghosh, Somsubhra and Sengupta, K. and Paul, Indranil},
  year = 2024,
  month = jan,
  journal = {Phys. Rev. B},
  volume = {109},
  number = {4},
  pages = {045145},
  publisher = {American Physical Society},
  doi = {10.1103/PhysRevB.109.045145}
}

@article{Shen2022CP,
  title = {Non-{H}ermitian Skin Clusters from Strong Interactions},
  author = {Shen, Ruizhe and Lee, Ching Hua},
  year = 2022,
  month = sep,
  journal = {Comm. Phys.},
  volume = {5},
  number = {1},
  pages = {238},
  publisher = {Nature Publishing Group},
  issn = {2399-3650},
  doi = {10.1038/s42005-022-01015-w},
  copyright = {2022 The Author(s)}
}

@article{Wang2025CPL,
  title = {Non-{H}ermitian Skin Effects in Fragmented {H}ilbert Spaces of One-Dimensional Fermionic Lattices},
  author = {Wang, Yi-An and Li, Linhu},
  year = 2025,
  month = mar,
  journal = {Chin. Phys. Lett.},
  volume = {42},
  number = {3},
  pages = {037301},
  publisher = {{Chinese Physical Society and IOP Publishing Ltd}},
  issn = {0256-307X},
  doi = {10.1088/0256-307X/42/3/037301}
}

@article{Li2023PRR,
  title = {Hilbert Space Fragmentation in Open Quantum Systems},
  author = {Li, Yahui and Sala, Pablo and Pollmann, Frank},
  year = 2023,
  month = dec,
  journal = {Phys. Rev. Research},
  volume = {5},
  number = {4},
  pages = {043239},
  publisher = {American Physical Society},
  doi = {10.1103/PhysRevResearch.5.043239}
}

@misc{Paszko2025,
  title = {Operator-Space Fragmentation and Integrability in {Pauli-Lindblad} Models},
  author = {Paszko, Dawid and Turner, Christopher J. and Rose, Dominic C. and Pal, Arijeet},
  year = 2025,
  month = jun,
  number = {arXiv:2506.16518},
  eprint = {2506.16518},
  primaryclass = {quant-ph},
  publisher = {arXiv},
  doi = {10.48550/arXiv.2506.16518},
  archiveprefix = {arXiv}
}

@article{SalaPRX2020,
  title = {Ergodicity Breaking Arising from {H}ilbert Space Fragmentation in Dipole-Conserving {H}amiltonians},
  author = {Sala, Pablo and Rakovszky, Tibor and Verresen, Ruben and Knap, Michael and Pollmann, Frank},
  journal = {Phys. Rev. X},
  volume = {10},
  pages = {011047},
  year = {2020},
  doi = {10.1103/PhysRevX.10.011047}
}

@article{Yang2020PRL,
  title = {Hilbert-Space Fragmentation from Strict Confinement},
  author = {Yang, Zhi-Cheng and Liu, Fangli and Gorshkov, Alexey V. and Iadecola, Thomas},
  year = 2020,
  month = may,
  journal = {Phys. Rev. Lett.},
  volume = {124},
  number = {20},
  pages = {207602},
  publisher = {American Physical Society},
  doi = {10.1103/PhysRevLett.124.207602}
}

@article{Pietracaprina2021,
  title = {Hilbert-Space Fragmentation, Multifractality, and Many-Body Localization},
  author = {Pietracaprina, Francesca and Laflorencie, Nicolas},
  year = 2021,
  month = dec,
  journal = {Ann. Phys.},
  series = {Special Issue on Localisation 2020},
  volume = {435},
  pages = {168502},
  issn = {0003-4916},
  doi = {10.1016/j.aop.2021.168502}
}

@article{MoudgalyaPRX2022,
  title = {Hilbert Space Fragmentation and Commutant Algebras},
  author = {Moudgalya, Sanjay and Motrunich, Olexei I. and Huse, David A.},
  journal = {Phys. Rev. X},
  volume = {12},
  pages = {011050},
  year = {2022},
  doi = {10.1103/PhysRevX.12.011050}
}

@article{Moudgalya2022,
  title = {Quantum Many-Body Scars and {H}ilbert Space Fragmentation: A Review of Exact Results},
  shorttitle = {Quantum Many-Body Scars and Hilbert Space Fragmentation},
  author = {Moudgalya, Sanjay and Bernevig, B Andrei and Regnault, Nicolas},
  year = 2022,
  month = jul,
  journal = {Rep. Prog. Phys.},
  volume = {85},
  number = {8},
  pages = {086501},
  publisher = {IOP Publishing},
  issn = {0034-4885},
  doi = {10.1088/1361-6633/ac73a0}
}

@article{Adler2024Nature,
  title = {Observation of {H}ilbert Space Fragmentation and Fractonic Excitations in 2{D}},
  author = {Adler, Daniel and Wei, David and Will, Melissa and Srakaew, Kritsana and Agrawal, Suchita and Weckesser, Pascal and Moessner, Roderich and Pollmann, Frank and Bloch, Immanuel and Zeiher, Johannes},
  year = 2024,
  month = dec,
  journal = {Nature},
  volume = {636},
  number = {8041},
  pages = {80--85},
  publisher = {Nature Publishing Group},
  issn = {1476-4687},
  doi = {10.1038/s41586-024-08188-0},
  copyright = {2024 The Author(s)}
}

@article{Wang2025PRXQ,
  title = {Exploring {H}ilbert-Space Fragmentation on a Superconducting Processor},
  author = {Wang, Yong-Yi and Shi, Yun-Hao and Sun, Zheng-Hang and Chen, Chi-Tong and Wang, Zheng-An and Zhao, Kui and Liu, Hao-Tian and Ma, Wei-Guo and Wang, Ziting and Li, Hao and Zhang, Jia-Chi and Liu, Yu and Deng, Cheng-Lin and Li, Tian-Ming and He, Yang and Liu, Zheng-He and Peng, Zhen-Yu and Song, Xiaohui and Xue, Guangming and Yu, Haifeng and Huang, Kaixuan and Xiang, Zhongcheng and Zheng, Dongning and Xu, Kai and Fan, Heng},
  year = 2025,
  month = feb,
  journal = {PRX Quantum},
  volume = {6},
  number = {1},
  pages = {010325},
  publisher = {American Physical Society},
  doi = {10.1103/PRXQuantum.6.010325}
}

@incollection{Moudgalya2020MV,
  title = {Thermalization and Its Absence within {K}rylov Subspaces of a Constrained {H}amiltonian},
  booktitle = {Memorial Volume for Shoucheng Zhang},
  author = {Moudgalya, Sanjay and Prem, Abhinav and Nandkishore, Rahul and Regnault, Nicolas and Bernevig, B. Andrei},
  year = 2020,
  month = nov,
  pages = {147--209},
  publisher = {WORLD SCIENTIFIC},
  doi = {10.1142/9789811231711_0009},
  isbn = {978-981-12-3170-4}
}

@article{WangC2023PRB,
  title = {Freezing Transition in the Particle-Conserving {E}ast Model},
  author = {Wang, Cheng and Zhi-Cheng Yang},
  year = 2023,
  journal = {Phys. Rev. B},
  volume = {108},
  pages={144308},
  number = {14},
  doi = {10.1103/PhysRevB.108.144308}
}

@article{Nakamura2012PRL,
  title = {Exactly Solvable Fermion Chain Describing a $\nu=1/3$ Fractional Quantum {H}all State},
  author = {Nakamura, Masaaki and Wang, Zheng-Yuan and Bergholtz, Emil J.},
  year = 2012,
  month = jul,
  journal = {Phys. Rev. Lett.},
  volume = {109},
  number = {1},
  pages = {016401},
  publisher = {American Physical Society},
  doi = {10.1103/PhysRevLett.109.016401}
}

@book{Tu2017Book,
  title = {Differential Geometry},
  author = {Tu, Loring W.},
  year = 2017,
  series = {Graduate Texts in Mathematics},
  volume = {275},
  publisher = {Springer International Publishing},
  address = {Cham},
  doi = {10.1007/978-3-319-55084-8},
  copyright = {http://www.springer.com/tdm},
  isbn = {978-3-319-55082-4 978-3-319-55084-8}
}

@article{Bergholtz2005PRL,
  title = {Half-Filled Lowest {L}andau Level on a Thin Torus},
  author = {Bergholtz, Emil J. and Karlhede, Anders},
  year = 2005,
  month = jan,
  journal = {Phys. Rev. Lett.},
  volume = {94},
  number = {2},
  pages = {026802},
  publisher = {American Physical Society},
  doi = {10.1103/PhysRevLett.94.026802}
}

@article{Wiersig2014PRL,
  title = {Enhancing the Sensitivity of Frequency and Energy Splitting Detection by Using Exceptional Points: Application to Microcavity Sensors for Single-Particle Detection},
  shorttitle = {Enhancing the Sensitivity of Frequency and Energy Splitting Detection by Using Exceptional Points},
  author = {Wiersig, Jan},
  year = 2014,
  month = may,
  journal = {Phys. Rev. Lett.},
  volume = {112},
  number = {20},
  pages = {203901},
  publisher = {American Physical Society},
  doi = {10.1103/PhysRevLett.112.203901}
}

@article{Hodaei2017Nature,
  title = {Enhanced Sensitivity at Higher-Order Exceptional Points},
  author = {Hodaei, Hossein and Hassan, Absar U. and Wittek, Steffen and {Garcia-Gracia}, Hipolito and {El-Ganainy}, Ramy and Christodoulides, Demetrios N. and Khajavikhan, Mercedeh},
  year = 2017,
  month = aug,
  journal = {Nature},
  volume = {548},
  number = {7666},
  pages = {187--191},
  publisher = {Nature Publishing Group},
  issn = {1476-4687},
  doi = {10.1038/nature23280},
  copyright = {2017 Macmillan Publishers Limited, part of Springer Nature. All rights reserved.}
}

@article{Chen2017Nature,
  title = {Exceptional Points Enhance Sensing in an Optical Microcavity},
  author = {Chen, Weijian and Kaya {\"O}zdemir, {\c S}ahin and Zhao, Guangming and Wiersig, Jan and Yang, Lan},
  year = 2017,
  month = aug,
  journal = {Nature},
  volume = {548},
  number = {7666},
  pages = {192--196},
  publisher = {Nature Publishing Group},
  issn = {1476-4687},
  doi = {10.1038/nature23281},
  copyright = {2017 Macmillan Publishers Limited, part of Springer Nature. All rights reserved.}
}

@article{Budich2020PRL,
  title = {Non-{H}ermitian Topological Sensors},
  author = {Budich, Jan Carl and Bergholtz, Emil J.},
  year = 2020,
  month = oct,
  journal = {Phys. Rev. Lett.},
  volume = {125},
  number = {18},
  pages = {180403},
  publisher = {American Physical Society},
  doi = {10.1103/PhysRevLett.125.180403}
}

@article{Mostafazadeh2002JMP,
  title = {Pseudo-{H}ermiticity versus {PT} Symmetry: The Necessary Condition for the Reality of the Spectrum of a Non-{H}ermitian {H}amiltonian},
  shorttitle = {Pseudo-Hermiticity versus PT Symmetry},
  author = {Mostafazadeh, Ali},
  year = 2002,
  month = jan,
  journal = {J. Math. Phys.},
  volume = {43},
  number = {1},
  pages = {205--214},
  issn = {0022-2488},
  doi = {10.1063/1.1418246}
}

@article{Bender2024RMP,
  title = {$\mathcal{PT}$-Symmetric Quantum Mechanics},
  author = {Bender, Carl M. and Hook, Daniel W.},
  year = 2024,
  month = oct,
  journal = {Rev. Mod. Phys.},
  volume = {96},
  number = {4},
  pages = {045002},
  publisher = {American Physical Society},
  doi = {10.1103/RevModPhys.96.045002}
}

@article{Bender1998PRL,
  title = {Real Spectra in Non-Hermitian Hamiltonians Having $\mathcal{PT}$ Symmetry},
  author = {Bender, Carl M. and Boettcher, Stefan},
  journal = {Phys. Rev. Lett.},
  volume = {80},
  issue = {24},
  pages = {5243--5246},
  numpages = {0},
  year = {1998},
  month = {Jun},
  publisher = {American Physical Society},
  doi = {10.1103/PhysRevLett.80.5243},
  url = {https://link.aps.org/doi/10.1103/PhysRevLett.80.5243}
}

@Article{SupMat,
  author={},
  year={},
  journal = {In the Supplemental Material, we give the details of the proof for the trivial holonomy of the minimal models except the most symmetric symmetry sectors, and the existence of emergent open boundaries over a range of fillings for the fermionic model. }
}

\section{End Matter}
\emph{Analogy to holonomy in differential geometry.}--- Here, we explain why we use the name \textit{discrete holonomy} for the Hilbert space structure, namely the analogy to the holonomy of continuous manifold in differential geometry.
In differential geometry, holonomy characterizes the transformation of the vector bundle under a parallel transport around a closed loop~\cite{Tu2017Book}.
In our case of Hilbert space, discrete lattice of many-body basis states, the vector bundle associated with each basis state is the set of integers $\mathbb{Z}$ and a $\hat{D}_i$ in the loop corresponds to an addition of $1$ and $\hat{D}^\dagger_i$ a $-1$ [see Fig.~\ref{fig_holo}(b)].
Starting from basis state $|f_j\rangle$ associated with an arbitrary integer $s_j\in \mathbb{Z}$ (a vector in the vector bundle), after a closed loop, $s_j$ remains its initial value, namely the vector bundle undergoes a trivial transformation, if the loop contains an equal number of $\hat{D}_i$ and $\hat{D}^\dagger_i$.
On the other hand, after a loop with unequal numbers of $\hat{D}_i$ and $\hat{D}^\dagger_i$, $s_j$ changes, corresponding to a nontrivial transformation of the vector bundle.
Therefore, in the Hilbert space, if all the loops contain equal numbers of $\hat{D}_i$ and $\hat{D}^\dagger_i$, we refer to the Hilbert space as hosting trivial discrete holonomy.
On the other hand, a loop with unequal numbers of $\hat{D}_i$ and $\hat{D}^\dagger_i$ defines the non-trivial holonomy.

\emph{Examples of non-trivial discrete holonomy.}--- Here we show other examples of Krylov sectors with non-trivial discrete holonomy for the minimal models. 

In the main text, we show, for the Krylov sectors of the fermionic model mappable to the non-interacting HN model~\cite{Gliozzi2024PRL}, that a Fock state can return to itself by $\hat{D}^\dagger$ only.
This is analogous to having PBCs in the Fock state space.
Such periodicity is not restricted to the sectors mappable to the HN model, e.g., $|10101100\rangle$ also gets back to itself by inwards hopping only.
Sectors without periodicity under unidirectional hopping can also exhibit non-trivial discrete holonomy.
For example, $|1001011100\rangle$ arrives at $|0101101100\rangle$ by two inwards hopping, and no inwards hopping can be applied further.
However, we obtain $|0101110010\rangle$ by a further outwards hopping, which is related to the initial state $|1001011100\rangle$ by a translation of two sites.
Thus by five such three-step actions, each composed of two inwards hopping followed by a outwards hopping, a closed loop is completed with the number of inwards hopping twice that of outwards ones. 

For the spin model, there are also Krylov sectors with PBCs in the basis state space. For example, $|0+-+-+-\rangle$ can return to itself by $\hat{D}$ only. 

\begin{figure}[t]
  \centering
  \includegraphics[width=0.48\textwidth]{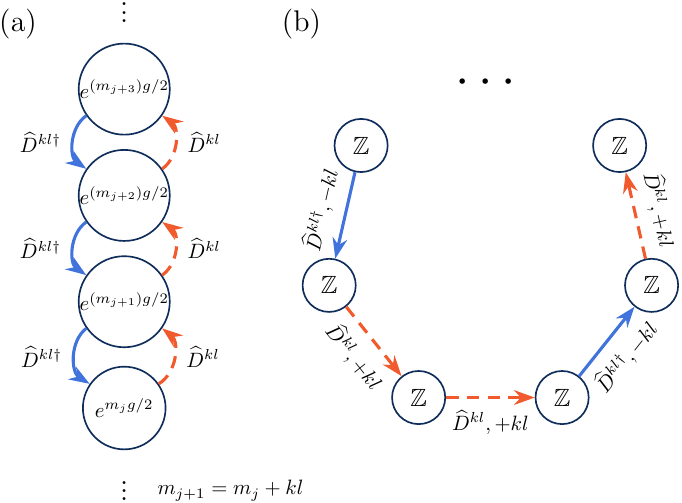}\hfill
  {\caption{\textbf{Schematic of the discrete holonomy for longer-range fermionic model}. (a) Schematic of the diagonal similarity transformation $\tilde{S}$ for the longer-range fermionic model (\ref{Long}). Each basis state is represented by a circle with the associated matrix element of $\tilde{S}$ labeled, with $\hat{D}^{kl}_i$ and $\hat{D}^{kl\dagger}_i$ corresponding to adding and subtracting a $g/2$ in the exponent. 
  (b) Schematic of a closed loop in the space of Fock states. Each state is associated with a $\mathbb{Z}$ bundle, with $\hat{D}^{kl}_i$ and $\hat{D}^{kl\dagger}_i$ corresponding to vector potential $A=+kl$ and $-kl$, respectively.
  }
  \label{fig_hololong}
  }
  
\end{figure}

\emph{Holonomy for longer-range models.}--- For the longer-range models (\ref{Long}) and (\ref{SpinLong}), the real spectra regimes are also ensured by trivial discrete holonomy of the Hilbert space.
The spin model (\ref{SpinLong}) shares the same definition of holonomy as the minimal models, as shown in Fig.~\ref{fig_holo}. 
Explicitly, $\hat{D}_i=S_i^+S_{i+1}^-S_{i+2}^-S_{i+3}^+$  and $\hat{D}_i^\dagger=S_i^-S_{i+1}^+S_{i+2}^+S_{i+3}^-$ correspond to $+1$ and $-1$ in the directed loop [see Fig.~\ref{fig_holo}(b)]  within the Hilbert space.
(Non-)Trivial holonomy is determined by the absence (presence) of a loop with unequal number of $\hat{D}_i$ and $\hat{D}_i^\dagger$.
The diagonal similarity transformation, which relates (\ref{SpinLong}) to its Hermitian counterpart, can be constructed as in Fig.~\ref{fig_holo}(a) in the case of trivial holonomy.

\begin{figure}[!t]
  \centering
  \includegraphics[width=0.48\textwidth]{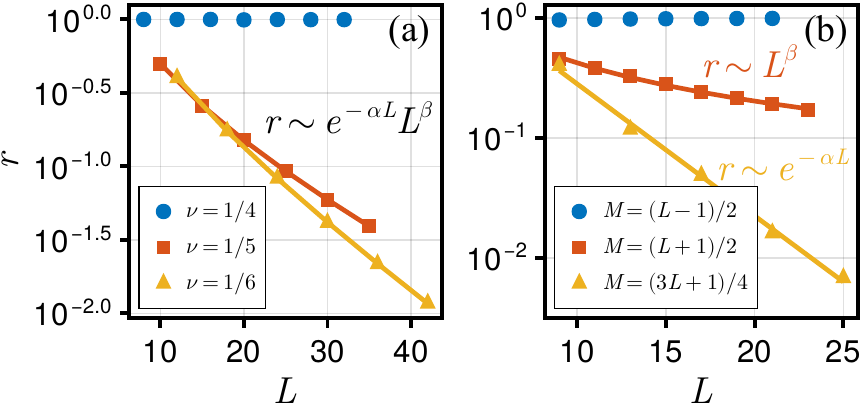}\hfill
  {\caption{\textbf{Strong and weak fragmentation}. Scaling of the ratio of the dimension of the maximal Krylov sector within a symmetry sector to the total dimension of the symmetry sector, $r=\mathrm{max}(\{\mathcal{D}_{K,i}\})/\mathcal{D}_K$, as a function of system size $L$ under PBCs. 
  For the fermionic (spin) model, besides the particle-number (magnetization), dipole moment $\hat{P}=\sum_i i c_i^\dagger c_i$ ($\hat{P}=\sum_i i S_i^z$) is also a conserved quantity.
  Therefore, the symmetry sector here is labeled by both particle-number (magnetization) and dipole moment. We take the symmetry sector with $P=0$ (mod $L$) as representative.
  (a) Fermionic model (\ref{Long}) with longer-range hopping. At $\nu=1/4$ (blue circle), $r\sim 1$ for all system sizes, demonstrating weak fragmentation. At $\nu=1/5$ (red square) and $1/6$ (yellow triangle), $r$ scales as $r\sim \exp{(-\alpha L)}L^{\beta}$ with $\alpha=0.05,\beta=-0.94$ for $\nu=1/5$ and $\alpha=0.08,\beta=-0.85$ for $\nu=1/6$, showing strong fragmentation. (b) Spin model (\ref{SpinLong}) with longer-range interaction. At $M=(L-1)/2$ (blue circle), $(L+1)/2$ (red square) and $(3L+1)/4$ (yellow triangle), $r$ scales as $r\sim 1$, $L^\beta$ with $\beta=-1.05$ and $\exp{(-\alpha L)}$ with $\alpha=0.25$, demonstrating weak, critical and strong fragmentation, respectively. 
  }
  \label{fig_strongweak}
  }
  
\end{figure}

For the fermionic model (\ref{Long}) with longer-range hopping, the holonomy structure is more complicated due to the multiple transport operators $\hat{D}_i^{kl}=c^\dagger_{i} c^\dagger_{i+k+l} c_{i+k} c_{i+l}$ with $0<l<k\leq k_{\mathrm{max}}$.
In the real spectra regime, we can construct a diagonal similarity transformation $\tilde{S}$ with matrix element $e^{m_jg/2}$ with integer $m_j$ for basis state $|f_j\rangle$, where $m_j-m_k=kl$ if $|f_k\rangle=\hat{D}^{kl}_i|f_j\rangle$ and $m_j-m_k=-kl$ if $|f_k\rangle=\hat{D}^{kl\dagger}_i|f_j\rangle$, as illustrated in Fig.~\ref{fig_hololong}(a). 
$\tilde{S}$ transforms (\ref{Long}) to its Hermitian counterpart, leading to the real spectra.
From the perspective of gauge potential, $\hat{D}^{kl}$ and $\hat{D}^{kl\dagger}$ correspond to $A=+kl$ and $-kl$, respectively.
A complete cycle across a directed loop within the Hilbert space, as shown in Fig.~\ref{fig_hololong}(b), contributes a flux $\phi=\sum_{\mathcal{C}} A$.
If there are no loops in the Hilbert space with nonzero $\phi$, the Hilbert space hosts trivial holonomy, enabling the construction of $\tilde{S}$ and ensuring real spectra.
Otherwise, the existence of a loop with nonzero $\phi$  determines the non-trivial holonomy, giving rise to complex spectra in principle.

\emph{Strong/Weak fragmentation.}--- In this section, we show the longer-range models (\ref{Long}) and (\ref{SpinLong}) exhibit strong and weak fragmentation at different filling factors or magnetization. Strong and weak Hilbert space fragmentation is classified by $r=\mathrm{max}(\{\mathcal{D}_{K,i}\})/\mathcal{D}_K$, where $\mathcal{D}_{K,i}$ is the dimension of Krylov sector $i$ in symmetry sector $K$ and $\mathcal{D}_K$ is the total dimension of symmetry sector $K$.
For strong fragmentation, $r$ decays to zero exponentially in the thermodynamic limit, i.e., $r\sim\exp(-\alpha L)$ and for weak fragmentation, $r\sim1$~\cite{SalaPRX2020}.

For the fermionic model (\ref{Long}), the ratio $r$ is always approximately equal to 1 at $\nu\geq1/4$ for all the system sizes $L$, with $\nu=1/4$ shown in Fig.~\ref{fig_strongweak}(a),  obviously demonstrating the weak fragmentation.
For $\nu\leq1/5$, $r$ scales as $r\sim \exp{(-\alpha L)}L^{\beta}$, with $\nu=1/5$ and 1/6 shown in Fig.~\ref{fig_strongweak}(a) for examples.
As exponential factor $\exp{(-\alpha L)}$ dominates over the polynomial factor $L^{\beta}$, the fragmentation at $\nu\leq1/5$ is identified as strong.
In the main text, we have shown the phase boundary between real and complex spectra regimes is $r\sim 1/5$ below half filling, which is close to the boundary between strong and weak fragmentation here, implying strong fragmentation favors real spectra.

For the spin model (\ref{SpinLong}) with odd $L$, the regime with $|M|\leq (L-1)/2$ exhibits weak fragmentation, demonstrated by $r\sim 1$ in Fig.~\ref{fig_strongweak}(b) with $M= (L-1)/2$ for example.
At $M= (L+1)/2$, $r$ scales as $r\sim L^\beta$, which is critical between strong and weak fragmentation~\cite{WangC2023PRB}.
For larger magnetization, we see $r\sim\exp{(-\alpha L)}$, with $M=(3L+1)/4$ shown in Fig.~\ref{fig_strongweak}(b), suggesting strong fragmentation.
We have found the minimal magnitude of $M$ with real spectra is $|M|=(L-1)/2$ for odd $L$ in the main text, at which the Hilbert space is weakly fragmented, suggesting trivial holonomy and real spectra can arise across strong and weak fragmentation.

\onecolumngrid

\section{Supplemental Material}


	\setcounter{equation}{0} \setcounter{figure}{0} \setcounter{table}{0} %
	\renewcommand{\theequation}{S\arabic{equation}} \renewcommand{\thefigure}{S%
		\arabic{figure}}

In the Supplemental Material, we give the details of the proof for the trivial holonomy of the minimal models except the most symmetric symmetry sectors, and the existence of emergent open boundaries over a range of fillings for the fermionic model.

\section{Details on the proof of the trivial discrete holonomy for minimal models other than the most symmetric symmetry sectors}
In this section, we give the details on the proof that the Hilbert space within any symmetry sector, except exact half-filling for the fermionic model and zero magnetization for the spin model, has trivial discrete holonomy. Namely, every closed loop in the Hilbert space contains an equal number of inwards and outwards hopping. 

We start from the fermionic model, where for each Fock state, we define height function
\begin{equation}
h_p=\sum_{i=1}^p(2n_i-1)-\frac{p(2n_e-L)}{L}
\end{equation}
as a function of the site index $p$, with $n_i\in\{1,0\}$ being the number of particles at site $i$.
By this construction, we have $h_{1}=h_{1+L}$, consistent with the PBCs.
The variance of $h_p$ is defined as 
\begin{equation}
V=\sum_{p=1}^L(h_p-\overline{h})^2,
\end{equation}
with $\overline{h}=\sum_{p=1}^Lh_p/L$.
Due to the PBCs, the way to label the sites from 1 to $L$ is not unique, but we can prove $V$ is independent of the initial site of the label.
To show this, we consider two sets of labels $\alpha,\beta$ with the initial sites differ by one site,
namely that $\beta=1$ and $\alpha=2$ refers to the same site.
We denote height functions under labels $\alpha,\beta$ as $h_p,\tilde{h}_p$, respectively, and  for $p=2,3,...,L$, $\tilde{h}_{p+1}$ and $h_p$ is defined on the same site while $\tilde{h}_{L}$ and $h_1$ are defined on the same site.
We have $\tilde{h}_{p+1}=h_p-(2n_1-1)+(2n_e-L)/L$ for $p=2,3,...,L$, with $n_1$ corresponding to $\alpha=1$, and $\tilde{h}_L=h_1+(2n_e-L)-(2n_1-1)-(L-1)(2n_e-L)/L=h_1-(2n_1-1)+(2n_e-L)/L$.
The change of the height function at every site is uniform, with the value $\Delta h_p=\tilde{h}_{p+1}-h_p=(2n_e-L)/L-(2n_1-1)$, thus the change of the average height function $\tilde{\overline{h}}-\overline{h}$ is also $\Delta_p$.
As a result, $\tilde{h}_{\tilde{p}}-\tilde{\overline{h}}=h_p-\overline{h}$ always holds for $\tilde{p}$ in $\beta$ and $p$ in $\alpha$ refer to the same site.
Therefore, for each action of $\hat{D}_i$ or $\hat{D}^\dagger_i$, we can choose the initial site in labeling, such that the range $i$ to $i+4$ does not cross the boundary (sites with labels $L$ to 1).
For the inward hopping $\hat{D}^\dagger_i$, the sites from $i$ to $i+4$ in the Fock state undergo $1001\rightarrow 0110$, and the changes of $h_p$ are $\Delta h_i=-2,\Delta h_{i+2}=2$ and $\Delta h_p=0$ for $p\neq i,i+2$. 
Obviously, $\overline{h}$ remains invariant and the change of the variance $V$ is $\Delta V_1=8+4(h_{i+2}-h_{i})=8(L-2n_e)/L$ obtained by simple algebra.
The opposite hopping $\hat{D}_i$ leads to a change of $-\Delta V_1$.
For a closed loop with $N_D$ outward hopping $\hat{D}_i$ and $N_{D^\dagger}$ inwards hopping $\hat{D}^\dagger_i$, the total change of $V$ is $\Delta V_{\mathrm{total}}=(N_{D^\dagger}-N_D)\Delta V_1$.
As $V$ is unique for every Fock state, $\Delta V_{\mathrm{total}}$ must be 0 for a closed loop, thus we have $N_D=N_{D^\dagger}$, namely trivial discrete holonomy, for $\nu\neq 1/2$ as $\Delta V_1\neq 0$ except at half-filling.
From the proof, we find the explicit form of the similarity transformation $\tilde{S}$ in general cases,  of which the matrix element corresponding to Fock state $f_j$ is $e^{[(V_j-V_0)/\Delta V_1 \rfloor g}$ where $V_0$ is a reference value defined per Krylov sector.

The trivial discrete holonomy of the spin model, except at zero magnetization, can be proved in the same spirit by defining $h_p=\sum_{i=1}^p S_i^z-pM/L$, where $S_i^z\in \{-1,0,1\}$ is the local spin at site $i$.
$\hat{D}_i$ renders a transition $(s_i,s_{i+1},s_{i+2})\rightarrow (s_i+1,s_{i+1}-2,s_{i+2}+1)$, leading to $\Delta h_i=1,\Delta h_{i+1}=-1$ and $\Delta h_{p}=0$ for other $p$.
Then $\Delta V_1=2(h_i-h_{i+1})+2=2-2s_{i+1}+2M/L=2M/L$, where the last equality is because $s_{i+1}=1$ for a valid $\hat{D}_i$ transition.
Similarly, the change of $V$ is $-2M/L$ for $\hat{D}^\dagger_i$.
Therefore, the change of $V$ after a closed loop is $\Delta V_{total}=(N_{D^\dagger}-N_D)2M/L$, leading to $N_{D^\dagger}=N_D$ because $\Delta V_{\mathrm{total}}=0$ due to the unique definition of $V$ for each basis state.

\section{Proof of the emergent open boundaries at $\nu<1/3$ and $\nu>2/3$}
In this section, we prove that for the minimal fermionic model (1) with (2) in the main text, there always exists emergent open boundaries. 
Namely, starting from any Fock state, there exists at least one site that cannot be flipped by any sequence of $\hat{D}_j$ and $\hat{D}_j^\dagger$.

For a Fock state, we denote $n_i=0,1$ as the particle number at the site with index $i$.
For a lattice chain under periodic boundary conditions, index $i$ and $i+tL$ with integer $t$ correspond to the same site.
We define $Z_i$ for site $i$, that $Z_{i}=Z_{i-1}+1$ for $n_i=0$ and $Z_{i}=Z_{i-1}-1$ for $n_i=1$, and we set $Z_0=0$.
Then we have $Z_{i+L}=Z_i+Q$, where $Q=L-2n_e$ with $n_e$ being the number of particles (1s in the Fock state representation).
We define a set 
\begin{equation}\label{InvaSet}
S=\{Z_j\ (\mathrm{mod\ }Q )|n_j=1\ \mathrm{for\ integer\ positive\ }j\}
\end{equation}
for each Fock state, and now we prove $S$ is invariant under the action of the Hamiltonian.
If occupation number from $i$ to $i+3$ is 1001, we have $Z_i=Z_{i-1}-1,Z_{i+1}=Z_{i-1},Z_{i+2}=Z_{i-1}+1,Z_{i+3}=Z_{i-1}$, and for 0110, we have $Z_i=Z_{i-1}+1,Z_{i+1}=Z_{i-1},Z_{i+2}=Z_{i-1}-1,Z_{i+3}=Z_{i-1}$.
We see for both 1001 and 0110, the set of $Z_j$ on sites with $n_j=1$ is $\{Z_{i-1}-1,Z_{i-1}\}$.
The only action of the Hamiltonian is $1001\leftrightarrow 0110$, thus $S$ is invariant under the action of any sequence of $\hat{D}_j$ and $\hat{D}_j^\dagger$.

$S$ contains at most $n_e$ elements. For $\nu<1/3$, $L>3n_e$, thus $Q>n_e$. Therefore, there must exist at least one positive integer $k\leq Q$ missing in $S$.
As $Z_j$ increases 1 for each site and $Q$ for a whole cycle, every positive integer is covered by $Z_i$ for $n_i=0$.
Therefore, there must exist site $s$ with $Z_s=k+1$ and $n_s=0$.
We have $Z_{s-1}=k$, arriving at $n_{s-1}=0$ due to the missing of $k$ in $S$.
We also have $n_{s+1}=0$, because otherwise we would get $Z_{s+1}=k$ with $n_{s+1}=1$, leading to contradiction. 
Therefore, we have three consecutive 0s for sites $s-1,s,s+1$, which is always invariant under the action of $\hat{D}_j$ and $\hat{D}_j^\dagger$, because any flip of there sites contradicts the invariance of $S$.

As $\nu>2/3$ is related to $\nu<1/3$ by particle-hole symmetry, we see there must also exist emergent open boundaries for $\nu>2/3$.

For filling factors $1/3\leq\nu\leq 2/3$, although the emergent open boundary picture fails in general, there also exist certain Krylov sectors with emergent open boundaries.
For example, at $\nu=1/3$, the number at every site in the Fock state $100100100...100$ can be flipped.
However, if we translate one particle by one site, e.g., $101\boxed{000}100...100$, the 0s in the box cannot be changed and an EOB arises. 
Note that three consecutive 0s or 1s do not always play the role of an emergent open boundary, instead it is dependent on the nearby sites.

\end{document}